\documentclass[aps,prb,superscriptaddress]{revtex4}
\usepackage{amsmath,amssymb}
\usepackage{bm}
\usepackage{graphicx}
\usepackage{epstopdf}
\newcommand{\beq}{\begin{equation}}
\newcommand{\eeq}{\end{equation}}
\begin{document}
\title{A unified quasiparticle approach to the theory of strongly correlated
electron liquids}
\author{V.~A.~Khodel}
\affiliation{McDonnell Center for the Space Sciences \&
Department of Physics, Washington University,
St.~Louis, MO 63130, USA}
\author{J.~W.~Clark}
\affiliation{McDonnell Center for the Space Sciences \&
Department of Physics, Washington University,
St.~Louis, MO 63130, USA}
\affiliation{Centro de Investiga\c{c}\~{a}o em Matem\'{a}tica
e Aplica\c{c}\~{o}es, University of Madeira, 9020-105
Funchal, Madeira, Portugal}
\author{M.~V.~Zverev}
\affiliation{National Research Centre Kurchatov
Institute, Moscow, 123182, Russia}
\affiliation{Moscow Institute of Physics and Technology,
Dolgoprudny, Moscow District 141700, Russia}

\begin{abstract}

Landau's quasiparticle formalism is generalized to describe a wide class of
strongly correlated Fermi systems, in addition to conventional Fermi liquids.
This class includes (i) so-called marginal exemplars and (ii) systems that
harbor interaction-driven flat bands, in both of which manifestations of
non-Fermi-liquid behavior are well documented. Specifically, the advent
of such flat bands is attributed to a spontaneous topological rearrangement
of the Landau state that supplements the conventional Landau quasiparticle
picture with a different set of quasiparticles, the so-called fermion
condensate, whose single-particle spectrum is dispersionless.  The celebrated
Landau-Luttinger theorem is extended to marginal Fermi liquids, in which
the density of the augmented quasiparticle system is shown to coincide
with the particle density. On the other hand, the total density of a system
hosting an interaction-driven flat band turns out to be the sum of the
densities of the two quasiparticle subsystems: the Landau-like component
and the fermion condensate.  We demonstrate that within the framework of
the scenario proposed,
a long-standing problem faced by theories of $D$-wave superconductivity in
cuprates, namely a consistent explanation of the so-called Uemera plot, can
be naturally resolved.
\end{abstract}
		
\maketitle

\noindent
{\bf 1. Introduction}	
\vskip 0.5 cm

The BCS paradigm, established more than half a century ago, has successfully
explained the phenomenon of superconductivity discovered by Kamerlingh Onnes
in 1911. This success rests upon (i) the Cooper scenario for electron pairing
in metals \cite{cooper,BCS,gor'kov} and (ii) the Landau Fermi-liquid (LFL)
theory. LFL theory is a specific quasiparticle formalism designed for the
normal state, in which these quasiparticles are presumed to be immortal,
with a zero-temperature momentum distribution $n_L(p) = \theta(p_F-p)$
identical to that of a perfect Fermi gas with Fermi momentum $p_F$
\cite{lan1,lan2,AGD}.  Electron-phonon interactions, responsible for Cooper
pairing in solids, are taken into account within the framework of a
nonperturbative Migdal-Eliashberg theory of superconductivity
\cite{migdal0,eliash}.

Subsequently, Larkin and Migdal successfully applied the BCS-LFL approach
to the description of superfluid 3D liquid $^3$He~\cite{LM}.  Their basic
result is this: In conventional Fermi liquids, where, in accord with the
Landau postulate, the damping of single-particle excitations is immaterial,
the London diamagnetic anomaly is expressed in terms of the carrier density
for an arbitrary interaction between particles. This implies that the
$T=0$ superfluid density coincides with the carrier density, irrespective
of the strength of correlations. This result is identical to the content
of the Leggett theorem, derived later by him~\cite{leggett0}.

The discovery in 1986 of exotic high-$T_c$ superconductivity in the
strongly-correlated electron systems of copper oxides by Bednorz and
M\"uller \cite{BM,leggett} dealt an irreparable blow to the BCS-LFL
picture, when applied to these strongly correlated electron systems.
BCS theory has proven to be inadequate for explanation of a wide range
of experimental data on superconductivity in the so-called strong-coupling
limit \cite{pjh2018_scl}, while LFL theory itself fails to explain the
normal-state resistivity $\rho(T)$ of exotic superconductors, often
called strange metals (SMs), in which~\cite{norman,ph2022}
\beq
\rho(T)-\rho_0=A_1T .
\label{nfl1}
\eeq
This behavior implies that in normal states of exotic superconductors, the
damping $\gamma(T)$ of single-particle excitations varies {\it linearly} with
$T$, in striking contrast with the LFL prediction $\gamma_{\rm LFL}(T)
\propto T^2$.

A marginal-Fermi-liquid (MFL) phenomenology introduced soon after the
Bednorz-M\"uller discovery was the first model capable of explaining the
observed non-Fermi-liquid (NFL) behavior (\ref{nfl1}). A key feature of
this phenomenology is the addition to the LFL self-energy $\Sigma_{\rm LFL}$
of a specific {\it non-analytic} term~\cite{MFL}
\beq
\Sigma_{\rm MFL}(\varepsilon,T)\propto\Big[\frac{2\varepsilon}{\pi}
\ln\Bigl(\frac{{\rm max} (|\varepsilon|,T)}{\varepsilon_c}\Bigr)
-i\,{\rm max}(|\varepsilon|,T)\Big] ,
\label{sigmfl}
\eeq
triggered in the original model by critical antiferromagnetic fluctuations
\cite{MFL1,MFL2,Varma2016,Varma2020}. In accord with Eq.~(\ref{sigmfl}), the
MFL damping of single-particle excitations does indeed change linearly with
energy $\varepsilon$, rendering explanation of the NFL behavior~(\ref{nfl1})
straightforward.  However, upon adoption of Eq.~({\ref{sigmfl}), the
quasiparticle weight $z$ vanishes, being traditionally determined by the
standard Dyson-like formula
\beq
z^{-1}=1-\Bigl(\frac{\partial{\rm Re} \Sigma(p_F,\varepsilon)}
{\partial\varepsilon}\Bigr)_{{\varepsilon=0}}.
\label{qw}
\eeq
Furthermore, since the first-derivative term appearing here diverges
logarithmically, higher terms being even more singular, it becomes
inevitable that the Landau quasiparticle formalism and MFL phenomenology
are {\it incompatible}, fostering intense debate about the ultimate nature
of basic degrees of freedom in SMs~\cite{ph2022}.

In the time that has elapsed since the introduction of MFL phenomenology, it
has become clear that MFL predictions are often at variance with experiment.
For example, since critical fluctuations soar upward in the vicinity of a
quantum critical point (QCP), so does the doping-dependent MFL coefficient
$A_1(x)$; yet away from it, the MFL resistivity demonstrates canonical
LFL $T^2$ behavior.  In cuprates, however, experiment shows quite different
behavior:  $A_1(x)$ spreads over an extended range of doping-$x$ variation,
coming to naught at a value $x_s$ where the SM and LFL states meet each
other \cite{norman,ph2022,hussey1,hussey2,bozovic1,bozovic2}.

Remarkably, in a number of heavy-fermion metals the linearity of $\rho(T)$
is accompanied by the most spectacular exhibition of NFL behavior, associated
with a gigantic enhancement of the low-temperature normal-state thermal
expansion coefficient $\alpha(T)$~\cite{steglich,steg2008,steg1,flouq}
beyond its LFL and MFL values, which are proportional to $T$ and $T\ln T$
respectively.  Substantial contradictions between predictions of the
MFL scenario and experiment arise in the SM superconductors, where
superconductivity emerges directly as an instability of the SM state
\cite{norman}.  Accordingly, as $x$ approaches $x_s$ from the SM side,
both the critical temperature $T_c(x)$ of termination of superconductivity
and the coefficient $A_1(x)$, the signature of the normal SM state,
decline and vanish in unison.
Moreover, the MFL phenomenology
fails to provide any elucidation of the famous Uemura plot of the Cooper
critical temperature $T_c(T_F)$ against the Fermi temperature
$T_F=p^2_F/2m_e$ (see below).

These profound failures render the MFL scenario for NFL behavior incomplete
and vulnerable, suggesting that the theory of NFL phenomena in strongly
correlated Fermi systems is hardly exhausted by this model alone.  Indeed, an
alternative approach to explanation of this behavior was proposed around the
same time \cite{ks,vol1,noz,physrep,vol1994,zphys,ktsn}.  In a new scenario,
NFL behavior is triggered by a specific topological rearrangement of the
Landau state, called {\it fermion condensation} (FC).  Within the FC scenario,
such a topological transition leads to formation of a fermion condensate,
namely an interaction-driven flat band identified as a dispersionless
spectrum $\epsilon({\bf p})$ in a momentum domain $\Omega$.  Accordingly,
the FC density of states $D(\varepsilon)$ acquires a $\delta$-like
contribution $\rho_{\rm FC}\delta(\varepsilon)$, in obvious analogy to the
singular Bose-liquid term $\rho_B\delta(\varepsilon)$ that was documented
experimentally in scattering of slow neutrons in liquid $^4$He.  This
correspondence warrants use of the terms fermion condensate and fermion
condensation in describing phenomena observed in strongly correlated Fermi
systems in which the density of states $D(\varepsilon)$ diverges in the
momentum region $\Omega$.

A decade ago, a precursor of such a topological transition (TT) associated
with the divergence of the density of states $D(\varepsilon)$ was observed
in the homogeneous two-dimensional electron liquid of MOSFETs~\cite{mosfet}.
Several years later, the presence of a fermion condensate itself was
revealed in ARPES studies performed in graphene intercalated by Gd
\cite{starke}. In addition, the merging of the spin- and valley-split Landau
levels at the chemical potential in a clean strongly interacting 2D electron
gas in silicon is confirmed by vailable experimental data~\cite{prl112}.

Given this background, an implicit question drives our immediate agenda:
Is it possible to upgrade the original LFL quasiparticle formalism to
equip it for description of a flattening of the single-particle spectrum
beyond the TT point, which gives rise to NFL behavior inherent in cuprates that
demonstrate exotic superconductivity?  As will be seen, the answer to
this question is affirmative.

\vskip 0.5 cm
\noindent
{\bf 2. Key features of the generic quasiparticle formalism}	
\vskip 0.5 cm

Extant inconsistencies between the LFL formalism and MFL phenomenology
have fostered a widespread opinion that in MFLs and other strongly
correlated Fermi systems, quasiparticles are simply nonexistent (see
e.g.~\cite{norman,senthil2021}). Intense debates and announcements on
this subject continue. Their apotheosis, formulated in Ref.~\cite{ph2022},
reads: ``One thing is clear in this regime. The particle picture breaks
down. ... The strange metal raises the distinct possibility that its
resolution must abandon the basic building blocks of quantum theory.''
However, as will be seen, there is no sound basis for such strong claims.
Indeed, as revealed in results from ARPES studies and available experimental
data on thermodynamic and kinetic properties, an inherent aspect of the
physics of metallic-state resides in the presence of the Fermi
surface (FS), inextricably related to the singular, pole-like structure of
the single-particle Green function $G({\bf p},\varepsilon)$.  Specifically,
the FS is comprised of the totality of points where the trajectory
$\varepsilon = \epsilon({\bf p})$ of a zero of the inverse propagator
$G^{-1}({\bf p},\varepsilon)$ crosses the energy line.  This striking feature,
present ubiquitously, applies not only to common metals, where LFL theory
has no contenders, but also to anisotropic, strange, and ``bad'' metals.
Variations in the structure of the distributions of zeros from one metal
to another, even of quite different types, turn out to be small (see Fig.~1).
Thus, as long as there exists a Fermi surface, {\it quasiparticles}
associated with singularities of the Green function $G({\bf p},\varepsilon)$,
remain fundamental degrees of freedom.

\begin{figure}[t]
\begin{center}
 \includegraphics[width=0.4\linewidth, height=0.22\linewidth] {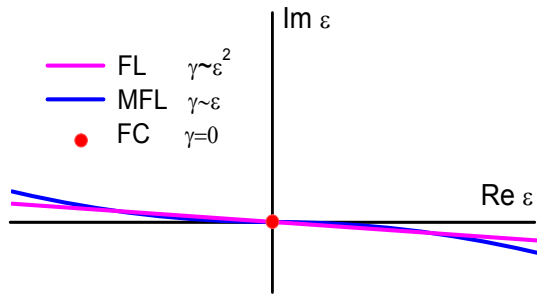}
\end{center}
\vskip -0.5 cm
\caption{Zeros of the inverse single-particle propagator
$G^{-1}(p_F,\varepsilon)$.  The magenta line corresponds to the Landau Fermi
liquid case with $\gamma_{\rm LFL}(\varepsilon)\propto \varepsilon^2$, whereas
the blue line tracks zeroes of the marginal Fermi-Liquid inverse propagator,
with $\gamma_{\rm MFL}(\varepsilon)\propto\varepsilon$. The red dot is related
to the fermion condensate.}
\label{fig:zeros}
\end{figure}

Any quasiparticle theory is designed to express quantitatively the low-$T$
characteristics of correlated Fermi systems in terms of those of a system
of quasiparticles, interacting through phenomenological pair forces.  In
conventional homogeneous Fermi liquids, the procedure for introducing
quasiparticles involves decomposition of the single-particle propagator
$G(p,\varepsilon) =(\varepsilon +\mu -p^2/2m-\Sigma(p,\varepsilon))^{-1}$,
with $\mu$ being the chemical potential and $\Sigma$ denoting the
self-energy, as an analytic function of energy, into the sum
\cite{lan2,AGD,pit,LM,migdal}
\beq
G(p,\varepsilon)\equiv zG^q( p,\varepsilon)+ G^r( p,\varepsilon).
\label{dec}
\eeq
The quasiparticle weight $z$ is calculated with the aid of the traditional
relation~(\ref{qw}), while $G^r$ represents the regular part of $G$.  The
inverse LFL quasiparticle propagator, determined as
$\bigl(G^q(p,\varepsilon)\bigr)^{-1}=\varepsilon-\epsilon(p)$,
coincides with that of a perfect Fermi gas of the same density $\rho$.
Near the Fermi surface,
the quasiparticle spectrum $\epsilon(p)$, possessing the single root
$p=p_F$, is expressed in terms of the self-energy $\Sigma$ through the
LFL formula
\beq
\epsilon(p)=z\frac{p_F(p-p_F)} {m} \Bigl(1+ \frac{m}{p_F}
\Bigl(\frac{\partial\Sigma(p,0)}
{\partial p}\Bigr)_{{p=p_F}}\Bigr),
\label{qps}
\eeq
where $m$ stands for the mass of a free particle.  This relation is usually
written in the Landau form
\beq
\epsilon(p)=v_F(p-p_F)
\label{lvf}
\eeq
in terms of the Fermi velocity $v_F=p_F/m^*$ and the effective mass
$m^*$.

An additional observation involves the damping $\gamma_{\rm LFL}(\varepsilon)$
of single-particle excitations, which in conventional Fermi liquids is
proportional to $\varepsilon^2$.  Since the ratio
$\gamma_{\rm LFL}(\varepsilon)/\varepsilon$ comes to naught, the role of
damping effects turns out to be immaterial, allowing LFL quasiparticles to
be treated as immortal.

\vskip 0.5 cm
\noindent
{\it 2.1. Introducing quasiparticles in marginal Fermi liquids}
\vskip 0.5 cm

The standard LFL procedure summarized above fails for MFLs, since the marginal
term $\Sigma_{\rm MFL}(\varepsilon)$ cannot be expanded in a Taylor series.
However, as we will readily demonstrate, the presence of this term is not an
unsurmountable obstacle to extension of the quasiparticle formalism to MFLs.
Indeed, the LFL strategy of introducing quasiparticles can be preserved,
provided the whole marginal contribution $\Sigma_{\rm MFL}(\varepsilon)$
is kept intact, while expanding in a Taylor series involves only the remaining
analytic part $\Sigma_a$ of $\Sigma_{\rm MFL}$.

It is worth noting that the idea of separating off non-analytic parts of the
self-energy $\Sigma$ that defy treatment within the standard Dyson procedure
is not new: it dates back to seminal articles by S.\ Belyaev and
L.\ Gor'kov~\cite{belyaev,gor'kov} that lay the foundation of theory of
Bose liquids and superconductivity (see below). In dealing with MFLs,
this strategy yields
\beq
G^{-1}( p\to p_F, \varepsilon\to 0)=
z^{-1}_a[\varepsilon-\epsilon(p)-\Sigma_m(\varepsilon)] .
\label{invg}
\eeq
Here the quasiparticle energy $\epsilon(p)$ is still given by Eq.~(\ref{qps}),
except that $\Sigma$ is replaced by $\Sigma_a$, while a remaining marginal
contribution takes the form
\beq
\Sigma_m(\varepsilon)=\gamma_1\varepsilon
\Bigl[\frac{2}{\pi}\ln(|\varepsilon|/\varepsilon_c)-i\Bigr],
\label{sigmmfl}
\eeq
with normalization chosen to match Eq.~(\ref{sigmfl}). In Eq.~(\ref{invg})
we have also introduced a refined {\it finite} quasiparticle weight
\beq
z^{-1}_a=1-\Bigl(\frac{\partial {\rm Re} \Sigma_a(p_F,\varepsilon)}
{\partial\varepsilon}\Bigr)_{{\varepsilon=0}} .
\label{za}
\eeq
With this upgrade, the revised formula for the inverse quasiparticle
propagator reads
\beq
\bigl(G^q(p,\varepsilon)\bigr)^{-1}\equiv z_aG^{-1}(p\to p_F,\varepsilon\to 0)
= \varepsilon-\epsilon(p)-\Sigma_m(\varepsilon) ,
\label{inp}
\eeq
revealing that at the FS, the first derivative of this function has a weak
logarithmic singularity.

\vskip 0.5 cm
\noindent
{\it 2.2. Quasiparticle structure of systems with interaction-driven flat bands}
\vskip 0.5 cm

Another option for the onset of NFL behavior, including the linear-in-$T$
variation of the resistivity $\rho(T)$, is associated with a spontaneous
violation of the topological stability of the original Landau
state~\cite{ks,vol1,prb2008,vol1994,ktsn,annals,m100,uspekhi,jltp2018}.
The quintessence of this phenomenon consists in the advent of the fermion
condensate, whose spectrum $\epsilon({\bf p})$ vanishes identically in
a compact region $\Omega$ of momentum space where the condensate propagator
takes the form~\cite{physrep}
\beq
G^q({\bf p},\varepsilon)=\frac{1-n_*({\bf p})}{\varepsilon
+i\delta} +\frac{n_*({\bf p})}{\varepsilon -i\delta},
\quad  {\bf p}\in \Omega .
\label{qfc}
\eeq
The condensate momentum distribution $0<n_*({\bf p})<1$, a continuous
function in momentum space, is normalized by the total  FC density
\beq
\rho_{\rm FC}=2\sum_{{\bf p}\in\Omega}n_*({\bf p}),
\label{fcrho}
\eeq
which in a homogeneous 3D matter becomes
\beq
\rho_{\rm FC}=2\int_{p_i}^{p_f} n_*(p)
\frac{d^3p}{(2\pi)^3} ,
\eeq
where $p_i$ and $p_f$ are lower and upper boundaries of condensate occupation.

\vskip 0.5 cm
\noindent
{\it 2.3. Mechanism of marginality of systems with interaction-driven flat bands.}
\vskip 0.5 cm

In systems hosting flat bands, the mechanism of marginality is known to be
associated with the scattering of LFL-like quasiparticles by the fermion
condensate~\cite{zphys,physrep,ktsn}, implying that the situation becomes
radically different from the that in the MFL case, where NFL behavior
is due to interaction between collective and single-particle degrees of
freedom \cite{MFL}.  To discuss this crucial distinction, we borrow a textbook
formula for evaluation of the damping rate $\gamma(\varepsilon)$.  In
symbolic notation adopted for brevity, it reads~\cite{AGD,pines}
\beq
\gamma(\varepsilon)=\Bigl(|\Gamma|^2 {\rm Im}G_R{\rm Im}G_R{\rm Im}G_R\Bigr).
\label{damp}
\eeq
Here $\Gamma$ is the scattering amplitude, while $G_R$ stands for the
retarded Green function.  Overwhelming contributions to the integral
involved come from the vicinity of the FS, which allows us to (i) neglect
contributions from regular parts of ${\rm Im}G_R$ and (ii) factor
$|\Gamma|^2$ out of this integral at the FS.  The NFL contribution
to the damping $\gamma$ comes from the condensate component of
${\rm Im}G_R({\bf p},\varepsilon)\propto n_*({\bf p})\delta(\varepsilon)$,
which, in accord with Eq.~(\ref{qfc}), differs from zero solely in the
FC momentum domain ${\bf p}\in\Omega$. In what follows we restrict
ourselves to realistic small values of the condensate factor
$\eta=\rho_{\rm FC}/\rho $, so that all the propagators but one
entering Eq.~(\ref{damp}) can be replaced by the corresponding LFL-like
expressions from Eq.~(\ref{inp}). Thereupon the structure of the
integrand becomes fully defined, rendering calculations routine
to yield
\cite{zphys}
\beq
\gamma_{\rm FC}(x,\varepsilon)\propto \eta(x)\,\varepsilon ,
\label{gamfc}
\eeq
which is proportional to the first power of energy, analogously to
the MFL result.
However, in the scenario for fermion condensation, the coefficient $A_1(x)$
that determines the damping of single-particle excitations has the form
\beq
A_1(x)\propto \eta(x)  .
\label{a1x}
\eeq
Presence of the condensate factor $\eta(x)$ in this formula makes a
profound difference: the magnitude of $A_1(x)$ turns out to grow
smoothly with variation of the difference $x_c-x\equiv x_s-x$
(see Fig.~3), while in the extant MFL phenomenology,
the NFL factor $A_1(x)$ is concentrated close to the QCP. We plan
to compare the predictions of the both approaches with available
experimental data~\cite{Tai0,Tai1,Tai2,Tai3} elsewhere

\vskip 0.5 cm
\noindent
{\it 2.4. Four topological critical points in overdoped cuprates}
\vskip 0.5 cm

We are now in position to comment on the onset of fermion condensation due
to violation of the topological stability of the Landau state in
strange metals,
taking as an illustration the family La$_{1-x}$Sr$_x$CO$_4$ of two-dimensional
cuprates, whose crystal lattice is square, as in many other high-$T_c$
superconductors.  Since the analysis of problems associated with presence of
the pseudogap is beyond the scope of this article, we address hereafter only
the overdoped domain of their phase diagrams.  At sufficiently large doping
values $x>0.3$, the original Landau state applies, and cuprate properties obey
LFL theory. However, as the $x$ value drops, there comes a point where
fermion condensation sets in, signaled by the appearance of four new
roots of the equation
\beq
\epsilon({\bf p}_c,x_c)=0,
\label{ecr}
\eeq
which, at a critical doping value $x_c$ in cuprates arise {\it simultaneously} in
momentum space at four {\it topological critical points} ${\bf p}_c$.
Such a specific degeneracy, due to the presence of a square crystal
lattice, is associated with four
saddle points, identified by momenta
${\bf p}_1=(0,\pi),{\bf p}_2 =(\pi,0)$ and ${\bf p}_1'
=(0,-\pi),{\bf p}_2' =(-\pi,0)$~\cite{vol1,ktsn}.
Evidently, in the MFL phenomenology, where collective degrees of freedom
govern the behavior, while the role of single-particle aspects is minimal,
such a degeneracy plays no significant role.

In concrete calculations of topological critical points, the following equivalent condition
turns out to be more convenient:
\beq
v({\bf p}_c,x_c)=|\nabla\epsilon({\bf p},x_c)|_{{\bf p}_c}=0,
\label{vcr}
\eeq
where $\nabla=\partial/\partial{\bf p}$ and $v({\bf p})$ is the quasiparticle
group velocity.  The underlying motivation is that calculations can then be
carried out on the disordered side of the transition point, where the
function $v({\bf p})$ is evaluated based on the Pitaevskii equation,
which has the standard
  form~\cite{AGD,pit}
\beq
  {\bf v}({\bf p}) = {\bf v}_0({\bf p})
+2\int f({\bf p},{\bf l})\nabla  n({\bf l}) \frac{d^2{\bf l}} {(2\pi)^2}
\label{nabf},
\eeq
where ${\bf v}_0({\bf p})=z{\cal T}^{\omega}({\bf p})/m_e$
is determined by the
$\omega$-limit of the vertex ${\cal T}^{\omega}({\bf p})=
\lim {\cal T}({\bf p},{\bf k}{\to}0,\omega{\to}0, kv_F/\omega{\ll}1)$
and a phenomenological Landau interaction function $f$ has been introduced.
The advantage of applying Eq.~(\ref{nabf}) in comparison with other approaches
lies in the fact that in this equation, the integration over momenta is
concentrated near the Fermi line,
and it is the structure of the interaction
amplitude in that region which determines the behavior of the relevant
phenomenological parameters. Results of numerical calculations based on
this equation, which will be analyzed in detail elsewhere, demonstrate that
the arrangement and configuration of the Fermi line is expressed profoundly
in the damping of single-particle excitations and hence in the NFL factor
$A_1(x)$.

\vskip 0.5 cm
\noindent {\bf 3. Upgrading the quasiparticle description for marginal
Fermi liquids}
\vskip 0.5 cm

The presence in the single-particle propagator of the small term
$\Sigma_m(\varepsilon)$, whose derivative diverges logarithmically at
$\varepsilon=0$, necessitates an upgrade not only of the extant procedure
for introducing quasiparticles, but also with respect to incorporation of
interaction effects. These tasks are performed along the same lines as in
LFL theory, with the aid of two fundamental relations of many-body theory
describing the response of Fermi systems to an external long-wavelength
longitudinal field $({\bf p}{\bf E})({\bf k}\to 0,\omega\to 0)$.  It is
essential that the form of the relations obtained remains independent of
the type of system under consideration. Correspondingly, there are two
different but complementary situations to consider.  In the first, where
${\bf k} = 0,\omega\neq 0$, one deals with a so-called diagonal perturbation
for which the field effect is {\it fictitious}, because the imposition of
the field results merely in shifting the energy argument in the propagator
$G({\bf p},\varepsilon)$.  A typical instance of such a perturbation is
associated with a variation of the Hamiltonian under a Galilean
transformation~\cite{pit,AGD,migdal}. Omitting details that the reader
can find in textbooks, we state here the final form of the appropriate
relation, taking 3D homogeneous matter as an example:
\beq
\frac{\partial G^{-1}({\bf p},\varepsilon)}{\partial\varepsilon}
{\bf v}_0 ={\bf v}_0 +
\int\!\!\!\!\int\!\!  {\cal U}({\bf p},\varepsilon,{\bf p}',\varepsilon')
\frac{\partial G^r({\bf p}',\varepsilon')}{\partial\varepsilon'}
{\bf v}'_0\frac{2d^3{\bf p}'\,d\varepsilon'}{(2\pi)^4i}.
\label{eqo}
\eeq
Here ${\bf v}_0={\bf p}/m_e$, while ${\cal U}$ denotes the block of
Feynman diagrams for the scattering amplitude irreducible in the
particle-hole channel, known to contain no quasiparticle contributions.

In the opposite limit $k/\omega\to\infty$, Eq.~(\ref{eqo}) is replaced by
a completely different relation stemming from gauge invariance,
\beq
- \nabla G^{-1}({\bf p},\varepsilon) = {\bf v}_0 + \int\!\!\!\!\int\!\!
{\cal U}({\bf p},\varepsilon,{\bf p}',\varepsilon') \nabla G({\bf p}',
\varepsilon') \frac{2d^3{\bf p}'\,d\varepsilon'}{(2\pi)^4i} .
\label{eqk}
\eeq

From this point, we employ for brevity a symbolic notation: round brackets
imply summation and integration over all intermediate variables,
supplemented by appropriate normalization factors. With this convention,
Eqs.~(\ref{eqo}) and (\ref{eqk}) take the forms
\begin{eqnarray}
\frac{\partial G^{-1}}{\partial\varepsilon}{\bf v}_0&=&{\bf v}_0+
\Bigl({\cal U}\frac{\partial G^r}{\partial\varepsilon}{\bf v}_0\Bigr),
\nonumber\\
- \nabla G^{-1}&=&{\bf v}_0 + \Bigl({\cal U}\nabla G\Bigr).
\label{nablas}
\end{eqnarray}
Following Landau's pioneering work~\cite{lan1}, we isolate quasiparticle
contributions to Eq.~(\ref{eqk}) by introducing a universal scattering
amplitude $\Gamma^{\omega}$ determined by
\beq
\Gamma^\omega={\cal U}+\Bigl(\Gamma^\omega B{\cal U}\Bigr)\equiv {\cal U}
+\Bigl({\cal U}B\Gamma^\omega\Bigr).
\label{rgamms}
\eeq
Here $B\equiv\lim_{\omega\to 0}G({\bf p},\varepsilon)
G({\bf p},\varepsilon+\omega)$ is, as traditionally, the regular part
of the product $GG$, devoid of quasiparticle contributions.

To continue, we multiply the first of Eqs.~(\ref{nablas}) from the left by
the product $\Gamma^\omega B$ and integrate.  Aided by Eq.~(\ref{rgamms}),
we then find
\beq
\frac{\partial G^{-1}({\bf p},\varepsilon)}{\partial\varepsilon}
{\bf p} =  {\bf p}+ \Bigl(\Gamma^\omega B{\bf p}\Bigr) .
\label{eq1}
\eeq
In gaining this result, we have accounted for the obvious cancellation
of two regular contributions.  The first is associated with
$\partial G^r/\partial\varepsilon$; the second, with
$-B\partial G^{-1}/\partial\varepsilon$.

Here we observe that Eq.~(\ref{eq1}) splits into two separate relations.  The
first is a standard equation containing the leading $\varepsilon$-independent
components of the self-energy and respective components of the interaction
function.  The second equation, more specific, is designed to generate
a logarithmic term $\propto \ln\varepsilon$ associated with the non-analytic
component $\Sigma_m$, whose magnitude is proportional to the small
condensate density $\rho_{\rm FC}$. If we are not interested in its evaluation,
the latter equation can simply be omitted.

In transformation of the second of Eqs.~(\ref{nablas}), analogous operations
are performed to produce
\beq
- \nabla G^{-1}({\bf p},\varepsilon\to 0)=\frac{\partial G^{-1}}
{\partial\varepsilon}{\bf v}_0- \Bigl(\Gamma^\omega A \nabla G^{-1}\Bigr),
\label{eq2}
\eeq
where
\beq
A=GG-B\equiv z^2_aG^qG^q,
\label{pla}
\eeq
yielding
\beq
A\nabla G^{-1}({\bf p},\varepsilon)= -z_a
\Bigl(\nabla G^q({\bf p},\varepsilon)
\Bigr).
\label{relaa}
\eeq

\vskip 0.5 cm
\noindent{\it 3.1. Persistence of the LL theorem}
\vskip 0.5 cm

We now turn to the relation between the particle and a quasiparticle
numbers, their equality in conventional Fermi liquids being the essence
of the celebrated Landau-Luttinger (LL) theorem~\cite{AGD,LW,l1960}.  We
begin by recasting the standard formula for evaluation of the electron
density $\rho$ in terms of the Fermi system's response to a weak longitudinal
electric field, already addressed briefly above,
\beq
\rho=\frac {N}{V}= - \frac{1}{3}{\rm Re}\int\!\!\!\!\int({\bf p}\nabla
G({\bf p},\varepsilon)) \frac{2d^3{\bf p}\,d\varepsilon}{(2\pi)^4i}.
\label{par}
\eeq
In symbolic notation, Eq.~(\ref{par}) is decomposed into the sum
$\rho=\rho_A+\rho_B$, with
\begin{eqnarray}
\rho_A&=&{\rm Re}\Bigl(({\bf p}\nabla G^{-1})A\Bigr),\nonumber\\
\rho_B&=&{\rm Re}\Bigl(({\bf p}\nabla G^{-1})B\Bigr).
\label{pars}
\end{eqnarray}
It is advantageous to begin the analysis with treatment of the second
of these integrals.  Substitution of Eq.~(\ref{eq2}) then yields
$\rho_B=\rho_B^{(1)}+\rho_B^{(2)}$, with
\beq
\rho_B^{(1)}=-{\rm Re}\Bigl(({\bf p}{\bf v}_0) B\frac{\partial G^{-1}}
{\partial\varepsilon} \Bigr)
\eeq
and
\beq
\rho_B^{(2)}={\rm Re}\Bigl( \Gamma^\omega B({\bf p} A \nabla G^{-1})\Bigr).
\eeq
We observe that the term $\Bigl(GG\partial G^{-1}/\partial\varepsilon\Bigr)
= -(\partial G/\partial\varepsilon)$ vanishes upon energy
integration, in turn triggering the vanishing of $\rho_B^{(1)}$.  We are
then left only with $\rho_B^{(2)}$, which, with the aid of Eq.~(\ref{eq1}),
is recast in the form
\beq
\rho_B^{(2)}={\rm Re}\Bigl(\frac{\partial G^{-1}}{\partial\varepsilon}({\bf p}
A \nabla G^{-1})\Bigr)-{\rm Re}\,({\bf p}A\nabla G^{-1}) .
\eeq
Its summation with $\rho_A$ given by the first of Eqs.~(\ref{pars}) produces
\beq
\rho= {\rm Re}\Bigl(\frac{\partial G^{-1}}{\partial\varepsilon}
({\bf p} A\nabla G^{-1})\Bigr),
\label{sumrho}
\eeq
which, with the aid of Eq.~(\ref{relaa}), yields the relation
\beq
\rho=-z_a{\rm Re}\Bigl(\frac{\partial G^{-1}}{\partial\varepsilon}({\bf p}
\nabla G^q({\bf p},\varepsilon))\Bigr) .
\label{fin0}
\eeq
At this point, we introduce a new integration variable
\beq
w=\varepsilon
-\gamma_1\varepsilon\,\Bigl[\frac{2}{\pi}\ln(\varepsilon/\varepsilon_c) -i\Bigr]
\eeq
in place of $\varepsilon$.  We then have
$dw=[\partial G^{-1}(\varepsilon) /\partial\varepsilon]d\varepsilon$
and
$G^q({\bf p},w) =\bigl(w-\epsilon({\bf p})\bigr)^{-1}$,
leading to the LFL integral
\beq
\rho=\int\!\!\!
\int{\rm Im}G^q({\bf p},w)\frac{2d^3{\bf p}\,dw}{(2\pi)^4}
=2\!\int\!\theta(p_F-p)\frac{d^3{\bf p}}{(2\pi)^3}=\frac{p^3_F}{3\pi^2}.
\label{flrho}
\eeq
Thus, the density of the refined quasiparticles of a marginal Fermi liquid
coincides with the particle density {\it provided} the Fermi momentum $p_F$
is determined, analogously to the LFL case, as the point where the singularity
of the single-particle Green function $G(p,\varepsilon)$ migrates into the
upper-half plane in $\varepsilon$.

\vskip 0.5 cm
\noindent{\it 3.2. MFL specific heat calculated in the unified quasiparticle
formalism}
\vskip 0.5 cm

With the above results in play, it is now instructive to determine whether
the refined quasiparticle formalism outlined above is capable of evaluation
of the MFL specific heat.  To this end, we invoke the generic
relation~\cite{AGD}
\beq
\frac{S(T\to 0)}{V}=-2\frac{\partial}{\partial T}
\Bigl[T\sum_{\omega_n}\int\ln{\cal G}({\bf p},\omega_n)
\frac{d^3{\bf p}}{(2\pi)^3}\Bigr],
\label{st}
\eeq
with $\omega_n=\pi T (2n+1)$.  In a first step, the temperature-dependent
Green function ${\cal G}({\bf p},\omega)$ is expressed straightforwardly
in terms of the retarded Green function $G_R$~\cite{AGD}:
\beq
\frac{S(T\to 0)}{V}= \frac{1}{T}\!\int\!\frac{2d^3{\bf p}}{(2\pi)^4i}
\!\!\int\limits_{-\infty}^{\infty}\!\!\varepsilon \Bigl(-\frac{\partial f(\varepsilon)}
{\partial \varepsilon}\Bigr) \ln\Bigl( \frac{G_R({\bf p},\varepsilon)}
{G_R^*({\bf p},\varepsilon)}\Bigr)d\varepsilon ,
\label{agd1}
\eeq
where $f(\varepsilon)=[1+\exp(\varepsilon/T)]^{-1}$.
In what follows, we retain only a logarithmically divergent part of
the derivative $dS(T\to 0)/dT$, which comes from the pole part $G^q_R$
entering the quantity $\ln\bigl(G_R^q({\bf p},\varepsilon)
/(G^q_R)^*({\bf p},\varepsilon)\bigr)$.  Simple algebra then leads
to $S=S_++S_-$, with~\cite{shag1}
\beq
S_{\mp}(T{\to}\,0)
\propto \!\int\limits_0^\infty
\!\frac{ z^2e^zdz}{(1+e^z)^2}\!\int\limits_0^\infty\! dw\tan^{-1}
\Bigl[\frac{\gamma_1}{\pm w{+}\gamma_1 \ln(\epsilon_c/T)}\Bigr] .
\eeq
Integrations over $w$ and $z$ are then separated to yield the
MFL result~\cite{MFL1,MFL2,MFL},
\beq
S(T\to 0)/T\propto C(T\to 0)/T  \propto \gamma_1\ln \frac{\varepsilon_c}{T}.
\label{smfl}
\eeq
Thus we conclude that there is no impassable abyss between conventional
Fermi liquids and marginal ones.  Rather, both may be properly described
within the framework of the quasiparticle formalism developed here.

\vskip 0.5 cm
\noindent{\bf 4. Quasiparticle formalism for systems hosting flat bands}
\vskip 0.5 cm

As seen above, Fermi systems harboring interaction-driven flat bands can
be treated as having two components, the first of which is the fermion
condensate and the second a set of LFL-like quasiparticles, a distinctive
feature being a linear-in-energy damping of single-particle excitations.
Expulsion of regular components of the single-particle propagators from
the generic expression for the total density $\rho$ can be performed
along the same lines as for MFLs in the derivation of Eq.~(\ref{sumrho}),
with the provision, in accord with Eq.~(\ref{fin0}), that the momentum
integration is to be performed over two different regions in which
the quasiparticle propagator $G^q$ takes different forms. After cumbersome
manipulations, we arrive at the formula
\beq
\rho=\rho_{\rm FC}-\frac{1}{3}{\rm Re}\int\limits_{{p\notin\Omega}}
\!\!\!\!\!
\int({\bf p}\nabla G({\bf p},\varepsilon))
\frac{2\,d^3{\bf p}\, d\varepsilon}{(2\pi)^4i},
\label{rhofcs}
\eeq
in which $\rho_{\rm FC}$ is given by Eq.\,(\ref{fcrho}).  Further integrations
are performed in analogy to those carried out in the preceding development
to yield
\beq
\rho=\rho_{\rm FC}+p^3_i/3\pi^2.
\label{rhofc}
\eeq
This key result provides a basis for the relation between particle and
quasiparticle densities in systems having a fermion condensate.

\vskip 0.5 cm
\noindent{\it 4.1. Entropy excess as a signature of systems harboring flat
bands}
\vskip 0.5 cm

Perhaps the most striking feature exhibited by normal states of systems
that harbor a flat band is the presence of a {\it finite} entropy excess,
which is evaluated with the aid of the combinatoric Landau-like
formula~\cite{physrep}
\beq
S_{\rm FC}=-2\sum_{{\bf p}\in\Omega}[ n_*({\bf p})\ln n_*({\bf p})
+(1-n_*({\bf p})) \ln(1- n_*({\bf p}))].
\label{exs}
\eeq
Remarkably, the behavior of this entropy excess -- {\it being proportional
to the FC density} $\rho_{\rm FC}$ -- is in stark contradiction to that of
the LFL and MFL entropies, which are proportional to $T$ and $T\ln T$
respectively.  Analogous unorthodox behavior is exhibited by the
coefficient of the thermal expansion
$\alpha(T)= l^{-1}(\partial l/\partial T) \equiv \partial S/\partial P$,
where $P$ is the pressure.  According to Eq.~(\ref{exs}), one has
\beq
\alpha_{\rm FC}(T\simeq T_c)\propto \rho_{\rm FC},
\label{afc}
\eeq
where $T_c$ is the critical temperature for termination of superconductivity.
The finite condensate density $\rho_{\rm FC}$ varies smoothly upon elevation
of $T$.

Within the LFL and MFL scenarios, modest values of $\alpha(T_c)$ are predicted,
being proportional to $T_c$ and $T_c\ln T_c$ respectively, whereas the
magnitude of the ratio $\alpha_{\rm FC}(T_c)/\alpha_{\rm LFL}(T_c)$ turns
out to be enormously enhanced, especially in heavy-fermion superconductors
that commonly have very low $T_c$ values.  Curiously, it is such a huge
enhancement effect that was uncovered 20 years ago in measurements performed
on the three-dimensional heavy-fermion superconductor CeCoIn$_5$, which has
a tiny critical temperature $T_c\simeq 2.3\,$K~\cite{steglich}. These
measurements demonstrated that the curve of $\alpha(T)$, leaving the
origin in accord with the Nernst decree $S(T=0)=0$, grows very rapidly
upon elevation of $T$, landing at $T\simeq T_c\ll T_D$ on a plateau whose
height $\simeq 4.0\times 10^{-6}\,$K$^{-1}$ exceeds the corresponding LFL
result $\alpha_{\rm LFL}(T\simeq T_c)$ by a factor of order $10^{3}$\,\,
\cite{steg1,steglich,steg2008}.  However, this most astonishing behavior,
often accompanied by exotic superconductivity (as addressed below) has
still not received due attention in the condensed-matter literature to
this day.

In the meantime, analogous enhancement factors have been repeatedly observed,
in particular quite recently, in the celebrated heavy-fermion metal
URu$_2$Si$_2$~\cite{flouq}, which undergoes a hidden second-order phase
transition at $T_c=17.5\,$K whose precise nature remains unknown.  In this
instance, the measured normal-state value of $\alpha(T\geq T_c) \simeq 5.0
\times 10^{-6}\,$K$^{-1}$ even exceeds that found in CeCoIn$_5$.  Despite
the almost tenfold superiority of the $T_c$ value inherent in URu$_2$Si$_2$,
the magnitude of the coefficient $\alpha(T\simeq T_c)$ still remains almost
$T$-independent, implying that this metal is not at all a LFL or MFL,
especially in view of the fact that the corresponding
value of the thermal expansion coefficient
is very anisotropic~\cite{flouq}.

Concluding the analysis of this subsection, we notice that within the
FC scenario and in consonance with Eqs.~(\ref{gamfc}) and (\ref{exs}), the
ratio $A_1(x)/\alpha(T_c,x)$ must be {\it doping-independent}:
\beq
\frac{A_1(x)}{\alpha(T_c,x)}= {\rm const.}
\label{rfb}
\eeq
Experimental verification of this scaling behavior involving thermodynamic
and kinetic characteristics would be of particular interest.

\vskip 0.5 cm
\noindent{\bf 5. Sperconducting state}
\vskip 0.5 cm

Adaptation of the foregoing theoretical development to the description
of superconducting systems that experience Cooper pairing is performed
within the framework of the BCS concept, but it now involves {\it two
different} single-particle Green functions $G_s$ and $F$. At $T=0$ these
obey the set of equations~\cite{AGD,gor'kov}
\begin{eqnarray}	
G_s({\bf p},\varepsilon)&=&\Bigl[G^{-1}({\bf p},\varepsilon)
+\Delta^2_0({\bf p})G(-{\bf p},-\varepsilon)\Bigr]^{-1} , \nonumber \\
F({\bf p},\varepsilon)&=&G(-{\bf p},-\varepsilon)\Delta_0({\bf p})
G_s({\bf p},\varepsilon).
\label{gorf}
\end{eqnarray}
Here the superconducting gap $\Delta_0$ is commonly assumed to be
$p-$ and
$\varepsilon-$independent, being determined from the Gor'kov equation
\beq
\Delta_0({\bf p};x)=-\int
\sum_{\bf p_1}
{\cal V}({\bf p},{\bf p}_1)
F({\bf p}_1,\varepsilon_1;x)\frac{d\varepsilon_1}{2\pi i} ,
\label{gapeq}
\eeq
where ${\cal V}$ is a block of Feynman diagrams irreducible in the
particle-particle channel.  The same relations are valid for the quasiparticle
components of $G_s^q$ and $F^q$ of the respective Green functions.  In
common metals without involvement of massive impurity-induced scattering,
the set of the quasiparticle Green functions has the standard form
\begin{eqnarray}
G_s^q({\bf p},\varepsilon)&=&\frac{u^2({\bf p})}{\varepsilon-E({\bf p})
+i\delta} +\frac{v^2({\bf p})}{\varepsilon+E({\bf p})-i\delta}, \nonumber \\
F^q({\bf p},\varepsilon)&=&\frac{\Delta_0({\bf p})}{(\varepsilon-E({\bf p})
+i\delta)(\varepsilon+E({\bf p})-i\delta)},
\label{polsup}
\end{eqnarray}
where $ u^2({\bf p})+v^2({\bf p})=1$ and
\beq
n({\bf p})=v^2({\bf p})=(E({\bf p})-\epsilon({\bf p}))/2E({\bf p})
\label{np}
\eeq
involves the Bogoliubov quasiparticle energy
\beq
E({\bf p})=\sqrt{\epsilon^2({\bf p})+\Delta^2_0({\bf p})},
\label{bog}
\eeq
with $\epsilon({\bf p})$ given by the FL relation $\epsilon({\bf p},n)
=\delta E/\delta n({\bf p})$.  Contributions of pairing terms also dependent
on the momentum distribution $n({\bf p})$ should be taken into account
to provide a specific $\Delta-$dependence of $\epsilon({\bf p})$~\cite{noz}.

The superconducting state is the vacuum with respect to the Bogolyubov
quasiparticles, which have momentum distribution $n_B({\bf p},T)
=[1+e^{E({\bf p})/T}]^{-1}$.  Insertion of this distribution into the
above formula (\ref{exs}) evidently yields $S(T=0)=0$, disposing of
any contradiction of the Nernst theorem.  As a result, the system becomes
superconducting with a critical temperature that substantially exceeds
the respective BCS value. Of course, Cooper pairing is not the unique
mechanism for ensuring recovery of the Nernst theorem down to zero
temperature.  For example, antiferromagnetism may replace pairing if
the Cooper scenario is ruled out for some reason, as in strong external
magnetic fields.  We expect to address these pertinent issues in a
sequel to this paper.

\vskip 0.5 cm
\noindent{\it 5.1. Equality of particle and quasiparticle numbers in
superfluid Fermi liquids}
\vskip 0.5 cm

In estimates of the gap value (e.g. in MFL phenomenology), the
simple consideration that the number of electrons participating in Cooper
pairing should be equal to the total particle number is commonly exploited.
But a natural issue then arises.  In Cooper systems, the standard Luttinger
method for proving the LL theorem becomes inappropriate.  However, in due
course we will find that in systems that experience Cooper pairing, the
equality between particle and quasiparticle numbers still persists.

To facilitate analysis, we first address conventional Fermi liquids such as
liquid $^3$He. Instead of Eq.~(\ref{par}), one now has
\beq
\rho = - \frac{1}{3}\int\!\!\!\!\int p_n\frac{\partial G_s({\bf p},
\varepsilon)} {\partial p_n} \frac{2d^3{\bf p}\,d\varepsilon}{(2\pi)^4i}
= \frac{1}{3} \int\!\!\!\!\int\!\! {\bf p}K_s({\bf p},\varepsilon) \nabla
G^{-1}({\bf p},\varepsilon) \frac{2 d^3{\bf p}\,d\varepsilon}{(2\pi)^4i} ,
\label{rhos}
\eeq
where
\beq
K_s({\bf p},\varepsilon)=\lim_{{\bf k}\to 0} [G_s({\bf p},\varepsilon)
G_s({\bf p} +{\bf k},\varepsilon)-F({\bf p},\varepsilon)
F({\bf p}+{\bf k},\varepsilon)] ,
\label{kss}
\eeq
and $p_n$ is a component of the momentum ${\bf p}$ normal to the Fermi surface.

In what follows, we employ results obtained in proof of the persistence of
the LL theorem in MFLs, decomposing $K_s$ into the sum of a quasiparticle
term $K^q_s$ and a regular remainder denoted $B_s$.

Insertion of the decomposition $K_s = K^q_s + B_s$ into Eq.~(\ref{rhos})
leads to
\begin{equation}
\rho=\rho_K+\rho_B,
\label{rhoLB}
\end{equation}
in which
\begin{eqnarray}
\rho_K&=&{\rm Re}\Bigl(({\bf p}\nabla G^{-1})K^q_s\Bigr),\nonumber\\
\rho_B&=&{\rm Re}\Bigl(({\bf p}\nabla G^{-1})B_s\Bigr).
\label{parc}
\end{eqnarray}
Hereafter we shall make use of two key relations, namely
\beq
\frac{\partial G^{-1}({\bf p},\varepsilon)}{\partial\varepsilon}
{\bf p} =  {\bf p}+ \Bigl(\Gamma^\omega B_s{\bf p}\Bigr),
\label{equs}
\eeq
and
\beq
- \nabla G^{-1}({\bf p},\varepsilon\to 0)=\frac{\partial G^{-1}}
{\partial\varepsilon}{\bf v}_0- \Bigl(\Gamma^\omega K^q_s \nabla G^{-1}\Bigr),
\label{eq2s}
\eeq
their derivations being analogous to those of Eq.~(\ref{eq1}) and
Eq.~(\ref{eq2}), respectively.

Combining this equation with the latter expression of Eq.~(\ref{rhos}) then
yields
$\rho_B = \rho_s^{(1)} + \rho_s^{(2)}$, in which the term
\beq
\rho_s^{(1)} = -{\rm Re}\Bigl(({\bf p}{\bf v}_0)B_s
\frac{\partial G^{-1}}{\partial\varepsilon}\Bigr)
=-\Bigl({\partial G_s}/{\partial\varepsilon}\Bigr)
\eeq
vanishes.  As a result, we are left with
\beq
\rho_B = {\rm Re}\Bigl (\Gamma^\omega B_s{\bf p}K^q_s \nabla G^{-1} \Bigr).
\eeq

These manipulations lead us finally to the result
\beq
\rho= \frac{1}{3}\int\!\!\!\!\int\!\!\frac{\partial G^{-1}({\bf p},
\varepsilon)}{\partial\varepsilon } p_n  K^q_s({\bf p},\varepsilon)
\frac{\partial G^{-1}({\bf p},\varepsilon)} {\partial p_n}
\frac{2d^3{\bf p}\,d\varepsilon}{(2\pi)^4i} .
\label{rhosq}
\eeq
Near the quasiparticle pole, one has $G_s({\bf p},\varepsilon)
=zG^q_s({\bf p},\varepsilon)$ and
\beq
K^q_s({\bf p},\varepsilon)=z^2 [G_s^q({\bf p},\varepsilon) G_s^q({\bf p},
\varepsilon) -F^q({\bf p},\varepsilon)F^q({\bf p},\varepsilon)],
\eeq
such that all the $z$-factors cancel to produce
\beq
\rho= - \frac{2}{3}\int\!\!\!\!\int p_n
\frac{\partial G_s^q({\bf p},\varepsilon)} {\partial p_n}
\frac{d^3{\bf p}\,d\varepsilon}{(2\pi)^4i}
=2\!\int \! v^2({\bf p})\frac{d^3{\bf p}}{(2\pi)^3} .
\label{lls}
\eeq
Accordingly, we have generalized the LL theorem, demonstrating coincidence
between the particle and quasiparticle densities in Cooper superconductors.
Extension to those electron systems that harbor interaction-driven flat
bands is straightforward.

\vskip 0.5 cm
\noindent{\bf 6. Applications to unconventional  high-$T_c$ superconductivity}
\vskip 0.5 cm

In this section, we address the gross structure of the $T-x$ cuprate phase
diagram, which demonstrates profound non-BCS behavior.  In doing so we focus
on an overdoped part of the phase diagram, because its underdoped part is
subject in full force to an enigmatic pseudogap phenomenon, whose nature
is beyond the scope of the present analysis.

\vskip 0.5 cm
\noindent{\it 6.1. Elucidation of non-BCS behavior of the Uemura plot}
\vskip 0.5 cm

The famous Uemura plot~\cite{uemura} showing critical temperature $T_c$
versus the Fermi energy $T_F=p^2_F/2m_e$ is one of the well-established
confirmations of the profound failure of the BCS-FL theory of BM
superconductors.  In Fig.~2, the experimental data are arranged
such that logarithms of critical temperatures $T_c$ are plotted on the
vertical axis and logarithms of the corresponding Fermi energies $T_F$,
on the horizontal axis.  As seen, most of the exotic superconductors
have $T_c(\rho)/T_F$ values around 0.01-0.05, whereas all of the
conventional BCS examples lie on the far right in the plot.  A more
vivid demonstration of the collapse of the mainstream BCS scenario and
non-BCS nature of BM superconductivity is hard to find.

To understand how the presence of interaction-driven flat bands restores
agreement between experiment and theory, it is helpful to recast the
gap equation~(\ref{gapeq}) in the standard simplified form:
\beq
\Delta_0({\bf p})=-\sum_{{\bf l}}{\cal V}({\bf p},{\bf l})
\frac{\Delta_0({\bf l})}{ 2E({\bf l})}.
\label{gapq}
\eeq
The quintessence of the FC scenario lies in the fact that in systems hosting
interaction-driven flat bands, overwhelming contributions to the r.h.s. of
Eq.~(\ref{gapq}) come from the condensate region ${\bf p}\in\Omega$ where the
ratio $\Delta_0/E({\bf p})$ remains of order unity, irrespective of how small
the gap value is. As a result, the condensate contributions to the r.h.s. of
Eq.~(\ref{gapq}) turn out to be of order unity, whereas the respective BCS
contributions vanish in the limit $\Delta\to 0$ as $\Delta_0\ln(\Omega_D/\Delta_0)$, justifying their neglect and leaving the result \cite{ks}
\beq
\frac{\Delta_0^{\rm FC}(x)}{T_F}\propto \frac{T_c^{\rm FC}(x)}{T_F}
\simeq 0.5g_{\rm FC}I_0
\label{gapfc}
\eeq
where
\beq
I_0(x)=\frac{2}{\rho}\sum_{{\bf p}\in\Omega}
\sqrt{n_*({\bf p})(1-n_*({\bf p}))}\propto\eta(x),
\label{int0}
\eeq
with $n_*(p)$ denoting the normal-state FC momentum distribution
and $\eta=\rho_{\rm FC}/\rho$ the dimensionless condensate density.
The quantity $g_{\rm FC}$ is the FC pairing constant, whose magnitude
can be markedly in excess of the BCS  constant $g$ (see below).

\begin{figure}[t]
\begin{center}
\includegraphics[width=0.4\linewidth, height=0.387\linewidth] {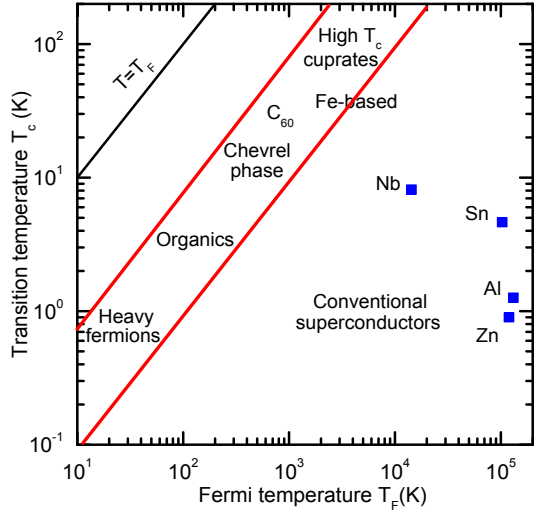}
\end{center}
\vskip -0.5 cm
\caption{Schematic Uemura plot showing the superconducting transition
temperature $T_c$ versus the Fermi temperature $T_F$~\cite{uemura, drozdov}.
Different families of unconventional superconductors are located between the two
red lines, while conventional superconductors lie far to the right.
}
\label{fig:Uemura_plot}
\end{figure}

At the optimal doping value $x_o$, Eq.~(\ref{gapfc}) becomes
\beq
T_c^{\rm max}/T_F\simeq 0.5g_{\rm FC}\eta_o,
\label{fctc}
\eeq
where $T_c^{\rm max}$ stands for the maximum value of the critical
temperature $T_c(x)$, while $\eta_o\equiv\eta(x_0)\simeq 0.1$ is
a dimensionless parameter.  Bearing in mind the enhancement of
the ratio $g_{\rm FC}/g$, we infer that within the framework of the
FC scenario, the Uemura plot can be properly explained, in contrast
to  results obtained in the MFL phenomenology.

\vskip 0.5 cm
\noindent{\it 6.2. Critical behavior of the ratio $T_c (x)/A_1(x)$}
\vskip 0.5 cm
  	
It is noteworthy that near critical doping value $x_c$ where the
superconducting dome terminates, two completely different quantities,
namely

(i) $A_1(x)$, a signature of the {\it normal} state that
identifies the linear-in-$T$ component of the resistivity $\rho(T,x)$
(whose NFL behavior has already been explained in the Section 2.2
(cf.\ Eq.~(\ref{a1x})), and

(ii) the critical temperature $T_c(x)$, a key feature of the
{\it superconducting} state,

\noindent
turn out to be proportional to the same quantity, namely the condensate
density $\eta(x)$, and both {\it come to naught in unison} at the boundary
between the SM and LFL states, due to vanishing of $\eta(x)$ at that point.
Thus we observe that these two ``birds,'' $A_1$ and $T_c$, can be
``killed'' with one ``stone,'' the role of the latter being played by
density of the fermion condensate (cf.\ Ref.~\cite{ph2022}).

\vskip 0.5 cm
\noindent{\it 6.3. Fermion condensate and $D$-wave superconductivity of
cuprates}
\vskip 0.5 cm

The $D-$wave character is another and perhaps more salient feature of BM
superconductivity of two-dimensional cuprates. To properly elucidate this
phenomenon within the framework of the FC scenario, we need to take
into account the fact that the symmetry of the crystal lattice promotes
simultaneous generation of four different condensate spots, each
associated with its own saddle point~\cite{vol1994,ktsn}. Thus the
domain of the momentum integration in Eq.~(\ref{gapq}) is divided
into four quite small condensate regions ${\bf p}\in \Omega_k$ where
the smooth functions
  ${\cal V}(\phi;{\bf l})$ and $\Delta({\bf l})$
remain practically unchanged and therefore can be factored out of the
integrands to yield \cite{JETPL2017}
\beq
\Delta(\varphi)=-\sum_k{\cal V}_k(\varphi) \frac{\Delta_k}{|\Delta|},
\label{delp}
\eeq
where $\Delta_k=\Delta({\bf p}_k)$ and ${\cal V}_k(\varphi)
={\cal V}(\varphi;{\bf p}_k)$. The four constants $\Delta_k$ are readily
determined based on their definition and Eq.~(\ref{delp}) to arrive at
the relation
\beq
\Delta_k=- \frac{I_0}{4} \sum_l{\cal V}_{kl}\frac{\Delta_l}{|\Delta|} ,
\label{delk}
\eeq
in which ${\cal V}_{11}={\cal V}(0,\pi;0,\pi), {\cal V}_{12}
={\cal V}(0,\pi;\pi, 0)$, etc.  Hereafter we also employ notation
${\cal V}^+={\cal V}_{12}\equiv{\cal V}_{1,2'} $.

The set of equations (\ref{delk}) has four different solutions.
The relation of our immediate interest is
\beq
\Delta_S(x)\simeq-\frac{\rho_{\rm FC}(x)}{8}\Bigl[{\cal V}_{11}
+ 2{\cal V}^+ +{\cal V}_{11'}],
\label{ds}
\eeq
which determines the gap $\Delta_S$ in the single-particle spectrum
due to $S-$wave pairing, provided the expression in the square brackets
has a negative sign; otherwise, $\Delta_S\equiv 0$. An analogous
statement is valid for the $D-$wave gap $\Delta_D$, given by
\beq
\Delta_D(x)\simeq-\frac{\rho_{\rm FC}(x)}{8}
\Bigl[{\cal V}_{11} -2{\cal V}^++{\cal V}_{11'}] .
\label{dd}
\eeq
Importantly, it should be noted that both the quantities $\Delta_S$
and $\Delta_D$, being proportional to the FC density $\rho_{\rm FC}$,
have nonzero values only on the ordered, topologically nontrivial side
of the Lifshitz transition.

To avoid unjustified complications, we restrict ourselves to a region
of the phase diagram where both $\Delta_D$ and $\Delta_S$ have
positive values. In this case, subtracting Eq.~(\ref{ds}) from
Eq.~(\ref{dd}) yields~\cite{arxiv2015}
\beq
\Delta_D(x)-\Delta_S(x)= 0.5\rho_{\rm FC}(x){\cal V}^+.
\label{dds}
\eeq
Thus, the outcome of competition between the $S-$wave and $D-$wave
pairing options depends solely on sign of the single matrix element
${\cal V}^+={\cal V}(0,\pi;\pi, 0)$, corresponding to the momentum
transfer $q\simeq \pi$ at which the electron-phonon exchange becomes
immaterial. Hence, in dealing with two-dimensional electron systems of cuprates the
sign of the difference~(\ref{dds}) and its
magnitude are primarily determined by the repulsive Coulomb part
of the matrix element ${\cal V}^+$ (cf. Eq.~(\ref{fctc})).

By way of illustration, Fig.~3 presents results of our estimation
of doping-dependence of the critical temperature $T_c(x)$~\cite{JETPL2017}.

\begin{figure}[t]
\begin{center}
\includegraphics[width=0.4\linewidth, height=0.387\linewidth]
{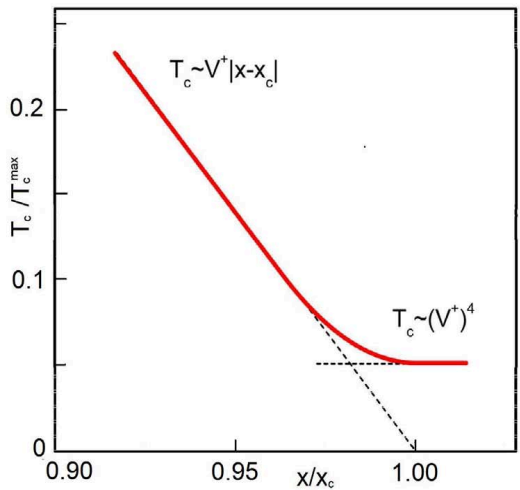}
\end{center}
\vskip -0.5 cm
\caption{Critical temperature $T_c$ versus doping $x$. As seen, at $x$
quite lower $x_c$, critical temperature $T_c(x)$ is proportional to
$x_c-x$, while in the vicinity of the topological critical point,
$T_c(x\simeq x_c)\propto ({\cal V}^+)^4$. This behavior agrees well
with the experimental data for La$_{2-x}$Sr$_x$CuO$_4$ obtained in
Ref.~\cite{ino}.}
\label{fig:T_c}
\end{figure}

\vskip 0.5 cm
\noindent {\bf 7. Conclusion}
\vskip 0.5 cm

In the present article, a unified quasiparticle theory appropriate
for the quantitative description of the challenging NFL behavior of
strongly correlated Fermi systems is presented.  The general framework
of this theory is in fact the same as in the Landau's conception, but
this extended formalism is free of shortcomings of the original
LFL picture, which is suitable only for conventional Fermi liquids
where damping of single-particle excitations is immaterial.

In a recent survey article~\cite{ph2022}, Phillips, Hussey, and
Abbamonte have raised several pertinent questions in addressing
the challenging problem of strange metallicity. In doing so, the authors
have proclaimed a breakdown of the quasiparticle pattern of phenomena
in condensed matter in the corresponding experimental domain, advocating
the introduction of non-quasiparticle states and/or anomalous dimensions.
Pending analysis of the results of quantitative application of our
extended quasiparticle theory to relevant experimental data on strange
metals, we are nevertheless prompted to contend that in dealing
conceptually with the theoretical picture advocated in Ref.~\cite{ph2022},
we have a typical scenario envisioned by Occam for the action of his
Razor, with the apt response:  ``When you hear hoofbeats, think horses,
not unicorns, not zebras." In the same spirit, working within an updated
quasiparticle approach, albeit mundane, we have developed a self-consistent
theory not only of strange metallicity, but indeed for diverse NFL phenomena
occurring in strongly correlated electron systems.

To be more specific, we summarize below the points where the Landau
quasiparticle formalism has been refined and extended, especially to cope
with the advent of topological disorder~\cite{prb2020}, the presence of a finite entropy
excess $S_*$ at extremely low $T \simeq T_c$, and the nature of the
Uemura plot $T_c(T_F)$.

(i) We have radically altered the original Dyson-like procedure for
introducing quasiparticles, in order to eliminate a fundamental drawback
of the Landau formalism that prevents it from yielding a correct description
of the observed non-Fermi-liquid behavior.

(ii) We have investigated the issue of equality between the densities of
particles and quasiparticles in electron systems that demonstrate
non-Fermi-liquid behavior.  This involves an extension of the celebrated
Landau-Luttinger theorem, its equality proof being applicable to conventional
Fermi liquids only.  In our treatment, equality has been proven for all
the systems under consideration, including their superconducting phases
as well, such a relation being hitherto unknown.

(iii) Within the framework of our approach, we have evaluated the
non-Fermi-liquid part of the electron heat capacity. The results obtained
coincide with those found in the marginal Fermi-liquid phenomenology. This
implies that our formalism is in fact applicable to the {\it quantitative}
description of these challenging quantum liquids.

(iv) On the basis of the fermion condensate picture we have explained the linear energy dependence of the quasiparticle damping rate responsible for the linear in temperature normal-state resistivity of strange metals.

(v) Within the framework of the fermion-condensate scenario, we have
explained the presence of a huge non-Fermi-liquid entropy excess
uncovered in measurements of the linear thermal expansion of strange
metals at temperatures comparable with the critical temperatures $T_c$
for termination of superconductivity~\cite{steglich, flouq}.

(vi) Within the framework of the same scenario, we have explained the
Uemura plot~\cite{uemura}, in which the non-BCS nature of Cooper pairing
in strange metals exhibits itself in full force.

(vii) On the same basis, we have properly explained the challenging doping
dependence of the critical temperatures $T_c(x)$ of LSCO compounds
measured by A.\ Ino et al. and reported in Ref.~\cite{ino}.

As the text of this article was being completed, we were made aware of a
recent essay by I.~Mazin~\cite{mazin}, in which the relevance of Occam's
razor was raised in connection with ideas popular in modern
condensed-matter physics. Our views happen to be very similar to
those espoused in this essay.

In conclusion, the authors express our gratitude to S.\ Kivelson,
P.\ Phillips, V.\ Shaginyan, C.\ Varma,  and G.\ Volovik for fruitful
discussions.

The work was carried out within the state assignment of NRC Kurchatov Institute.

\end{document}